# Gated Conductance of Thin Indium Tin Oxide — The Simplest Transistor

Jie Jiang[1,2], Qing Wan[1,a)], Jia Sun[1], Wei Dou[1], and Qing Zhang[2,b)]

1) Ningbo Institute of Materials Technology and Engineering, Chinese Academy of Sciences, Ningbo 315201, People's Republic of China.
2) NOVITAS, Nanoelectronics Centre of Excellence, School of Electrical and Electronic Engineering, Nanyang Technological University, 639798, Singapore.

**Abstract**: Transistors are the fundamental building block of modern electronic devices. So far, all transistors are based on various types of semiconductor junctions. The most common bipolar-junction transistors and metal-oxide-semiconductor field-effect transistors contain p-n junctions to control the current, depending on applied biases across the junctions. Thin-film transistors need metal-semiconductor junctions for injecting and extracting electrons from their channels. Here, by coating a heavily-doped thin indium-tin-oxide (ITO) film through a shadow mask onto a biopolymer chitosan/ITO/glass substrate, we can have a high-performance junctionless transparent organic-inorganic hybrid thin film transistor. This could be the simplest transistor in the world, to our knowledge, not only in its structure, but also its fabrication process. In addition, the device performance is found to be greatly enhanced using a reinforced chitosan/$SiO_2$ hybrid bilayer dielectric stack. Our results clearly show that this architecture can lead to a new class of low-cost transistors.

**Key words:** Chitosan-based biopolymer electrolyte; Junctionless thin-film transistor; Organic-inorganic hybrid electronics; Electric-double-layer.

a) **Email:** wanqing@nimte.ac.cn
b) **Email:** eqzhang@ntu.edu.sg

Field-effect transistors (FETs) are a broad type of semiconductor devices which modulate the current flow through applied gate voltages. They are the fundamental building blocks in many electronic products. In all FETs, the channels are lightly doped and device fabrication is always based on the formation of junctions between the channels and source/drain contacts[1]. Bipolar-junction transistors (BJTs) and inversion-mode metal-oxide-semiconductor field-effect transistors (MOSFETs) contain p-n junctions formed by either implantation or diffusion of donors or acceptors into a semiconductor substrate. The p-n junctions modulate the current according to the potential drop across them. It should point out that, in principle, the simplest FET can be constructed using an ultra-thin semiconductor in which charge carriers can be completely depleted by an electrical field applied through its gate dielectric layer/gate electrode. This conceptual FET without involving any junctions was patented by Lilienfeld in 1925 (ref. 2). However, as thin film technology was not available till recent decades, the simplest FET has not been fabricated successfully yet. Instead, to fully modify the concentration of charge carriers in the lightly-doped channel, metal-semiconductor junctions are usually employed for injecting and extracting electrons from the channels in thin-film transistors (TFTs). With rapid development of semiconductor technologies, a variety of transistors have been invented based on different structures and junction designs. Transistors with junctions are fundamental elements in many electronic products and they have played an important role in the evolution of electronic industry. In contrast, the concept of simple Lilienfeld's FET fades out of research attention.

As the p-n junction formation is formed through diffusion or implantation of dopant atoms into a semiconductor material, involvement of the junctions into transistors requires much more complicated design and fabrication processes than what the Lilienfeld FET would need. This no doubt increases the fabrication cost and bring a big technical challenge of precisely controlling the space profile of dopant atoms when the distance between two adjacent junctions is less than 10 nm in state-of-the-art integrated circuits. To break through the current process limitation and further reduce the fabrication cost, junctionless nanowire transistors have recently been demonstrated[3,4]. Based on a heavily doping Si-nanowire on a Si-on-insulator (SOI) substrate, Colinge et al built their transistor on silicon nanowires which had homogeneous doping polarity and uniform doping concentration across the channel without any junction formation two years ago[3,4]. Although the junction formation was not needed, the device fabrication still required rigorous multigate Si-nanowire patterned by expensive electron-beam lithography on SOI substrates. Thereafter, a lot of efforts have been paid for further relaxing the junctionless process by using other approaches[5-7]. Gundapaneni *et al* proposed a sort of bulk planar junctionless transistors by introducing an additional junction in the vertical direction for isolation purposes[5]. Lin *et al* reported junctionless poly-Si TFTs by using two-step-deposited channel and source/drain with an ultrathin dielectric[6]. A flash memory based on a gate-all-around Si-nanowire junctionless transistor was also demonstrated by Choi and his co-workers[7]. However, these junctionless transistors are silicon based devices and strongly rely on the stringent requirements, such as SOI wafers, ultrathin gate dielectric, precise multistep photolithography, etc. In other words, fabrication of these junctionless FETs is still rather challenging.

Recently, transparent oxide-based electronic devices are expected to meet emerging technological demands where silicon-based electronics can not provide a solution[8-10]. Oxide-based semiconductors have thus attracted much attention due to their high transparency and large carrier mobilities even in an amorphous structure deposited at room temperature[8-10]. Among the various oxide semiconductors, indium-tin-oxide(ITO) is a very promising material. A typical ITO film is an n-type degenerate semiconductor[11] with a wide bandgap of 3.8~4.0 eV and it has been widely investigated as a transparent conducting material in various optoelectronic applications such as flat-panel displays, solar cells, and organic light-emitting diodes, etc[12]. In contrast, little is known for its active electronic applications as a semiconductor material.

Another challenge of having high performance junctionless FETs originates from the quality of the gate dielectric. From the device physics perspective, successful operations of junctionless FETs depend on whether the charge carriers can be fully depleted in their channels in response to the applied gate voltages[6]. To achieve an efficient gating effect, a large gate capacitance is desirable for enhancing the electrostatic coupling between the gate and channel. Reducing the dielectric thickness is the most straightforward solution as the gate capacitance is proportional to the permittivity of the dielectric material, but inversely proportional to the thickness of the dielectric layer. When the thickness of $SiO_2$ or $Si_3N_4$, very commonly used dielectric materials, is reduced to ~ 10 nm or less, the leakage current through the gate layer would severely affect the device operations[13,14]. Another attractive strategy is to employ high-permittivity(high-$k$) materials such as metal oxide. Unfortunately, high-$k$ dielectric films are typically grown/deposited at high temperatures to ensure highly insulating characteristics[15,16]. To address the above issues, recently, a new type of dielectric, organic solid electrolytes/ionic liquids, has been reported as an fascinating approach to increase the capacitance due to large electric-double-layer(EDL) capacitance($>1\mu F/cm^2$) at dielectric/channel interface[17-22]. For example, Frisbie and Berggren demonstrated polyelectrolyte/ion gel as an effective gate insulator with a large capacitance($>1\mu F/cm^2$) for low-voltage TFTs applications[18-21]. High density carrier accumulation($>10^{14}/cm^2$) in ZnO-based FETs was also reported by Iwasa *et al* using ionic liquids as the gate dielectric[22]. In addition, we have also demonstrated that the microporous $SiO_2$ dielectric is an effective solid electrolyte with high gate control ability for several specific FET operations such as vertical[23], in-plane gate[24,25], and dual-gate TFTs[26]. Although these pioneering reports have attracted a lot attention, the exploitation of new dielectric materials is still of great significance.

Chitosan is a cationic biopolymer obtained from deacetylation of chitin, which is the second abundant natural polysaccharide on earth after cellulose[27,28]. It has found many applications in biotechnology, biomedicine, and food-packaging, due to its biocompatibility, biodegradability, non-toxicity, and excellent film-forming ability[27,28]. However, to our knowledge, very little attention has been paid to their potential dielectric application yet.

In this paper, we demonstrate two advances in tackling the challenges addressed above. We have successfully fabricated junctionless transparent TFTs gated through chitosan-based solid-biopolymer electrolyte. The devices have the following major features. (i) Extremely simple device structure: an ultra-thin highly-doped ITO is deposited though a shadow mask on a dielectric layer coated conducting substrate in just one step. This is the simplest transistor in the world in our knowledge. (ii) Novel dielectric material: chitosan is employed as a new low-cost solid-biopolymer-electrolyte dielectric to achieve high gate control ability and low threshold gate voltage. (iii) High performance organic-inorganic hybrid gate stack: a $SiO_2$ film(~5nm)/chitosan hybrid bilayer is found to be an efficient way to improve the stability and performance of the devices. (iv) The devices can be fabricated at room temperature.

Our junctionless TFTs are fabricated on conducting ITO glass substrates at room temperature, as shown in Fig.1. Firstly, chitosan solution (2 wt% in acetic acid) is drop-casted onto ITO-based substrates and dried in air to form a homogeneous chitosan film (thickness: ~6 μm). Secondly, an ultra-thin $SiO_2$ layer(~5nm) is deposited on the chitosan film using plasma-enhanced chemical vapor deposition (PECVD) method. For a comparison, the control samples are not coated with the thin $SiO_2$ layer. Thirdly, ITO films with different thickness(10nm, 20nm, 30nm, 60nm) are, respectively, deposited on the chitosan/$SiO_2$ dielectric using radio-frequency magnetron sputtering of an ITO target (90 wt% $In_2O_3$ and 10 wt% $SnO_2$) under a power of 100 W, a working pressure of 0.5 Pa and an Ar flow rate (14 sccm). The ITO films are patterned through a nickel shadow mask with the dimension of 150 μm×1000 μm. The capacitance-frequency measurements are performed using the WK 6500B precision impedance analyzer. The transfer/output characteristics of the devices are measured with a Keithley 4200 semiconductor parameter analyzer at room temperature in dark.

The specific gate capacitances ($C_i$) of the electrolyte dielectrics without and with the $SiO_2$ layer, using the sandwiched structure of ITO/chitosan(/$SiO_2$)/ITO, are shown in Fig.2(a) in the frequency range (20Hz-1M Hz). The inset displays the molecular structure of chitosan. For pure chitosan dielectric (black curve), $C_i$ increases with decreasing frequency and reaches to a maximum of 1.4 μF/cm$^2$ at 20 Hz. The strong frequency dependence of capacitance can be explained by the electrolyte behavior due to the presence of mobile ions in chitosan-based biopolymer[29,30]. Here, since acetic acid is used as the solvent, the chitosan-based biopolymer can exhibit protonated electrolyte behavior due to amino-group protonation ($-NH_2 + H^+ \Leftrightarrow -NH_3^+$). When an extrinsic electric field is applied, some of protonated amino groups will be deprotonated. Then, the protons can transport through hopping from one oxygen atom to another through hydrogen bonds, as shown in Fig.2(b). This mechanism is usually termed as the Grotthuss mechanism[29]. Finally, mobile protons($H^+$) and acetates($CH_3COO^-$) will move oppositely in response to the applied electric field and accumulate at the dielectric/electrode interfaces to form EDL capacitors. While at the high-frequency region, the low ion mobility limits the response time, and only a small number of ions can accumulate at the interface, leading to a low capacitance value[30].

This frequency dependence of the capacitance is consistent with the other results from solid electrolytes/ionic liquids[18-22,29,30]. Compared to a pure chitosan dielectric, the reinforced chitosan/SiO$_2$ hybrid bilayer dielectric (red curve) exhibits a capacitance value of 1.0 μF/cm$^2$ at 20 Hz. In addition, the hybrid dielectric has a better insulating property with a leakage current of ~0.6 nA at $V_G$=2.0V, a factor of three smaller than that of pure chitosan dielectric, see Fig.2(c).

The transfer and output characteristics of the TFTs gated through pure chitosan dielectric are shown in Fig 3. The transfer curves of the TFTs with different top ITO thicknesses ($t_{ITO}$) at different $V_{DS}$ (2.0V, 0.1V) are given in Figure 3(a). With $t_{ITO}$ =60 nm, the gate voltage ($V_{GS}$) has no a significant modulation effect on the drain current ($I_{DS}$). For $t_{ITO}$ =30 nm, the field-effect modulation is showing up, but very weak. However, under $t_{ITO}$ =20 nm, the $I_{DS}$ is found to be strongly dependent on $V_{GS}$. The subthreshold swing($S$) and on-off ratio($I_{on/off}$) are found to be 352 mV/dec and 4.2×10$^4$, respectively. It is noted that, when $t_{ITO}$ is further decreased to ~10nm, the TFTs ($V_{DS}$=2.0V) exhibit a much better FET performance with an $S$ value of 110mV/dec and $I_{on/off}$ of 1.2×10$^6$. Figure 3(b) shows the output curves of the TFT with $t_{ITO}$=10 nm. The $V_{GS}$ is varied from -0.2 to 1.0 V in 0.2V steps. The device exhibits clear current saturation behaviors at high $V_{DS}$. The linear characteristics without current crowding at low $V_{DS}$ suggest that the device has nice Ohmic contacts.

In a traditional inversion-mode or accumulation-mode MOSFET, the conducting path is close to the dielectric/channel interface due to the confinement of electric field originating from the gate electrode. However, for our junctionless ITO-TFTs, the operating mechanism is different from that of traditional devices and can be described as follows. In the subthreshold region (under a negative gate bias), the heavily-doped ITO channel is fully depleted under a negative gate voltage through the large EDL capacitance($>1\mu F/cm^2$), leading to an upward band bending of the ITO channel, as shown in Fig.4(a). With increasing the gate voltage positvely, the electric field in the channel reduces until a neutral region appears on top of the ITO channel, as shown in Fig.4(b). At this point, we should note that the current starts to flow through the neutral heavily-doped ITO channel. By further increasing the gate voltage, the depletion width decreases until a full neutral ITO is restored in the whole channel. This occurs when the gate voltage equals the flat-band voltage, as shown in Fig.4(c). As highly doping ITO is also employed as the gate metal in these devices, the flat-band voltage should be corresponding to $V_G \approx 0$. At the onset of this condition, the current reaches to a saturation region. Thereafter, by increasing the gate voltage further, electrons will accumulate on the inner surfaces of the channel, as shown in Fig.4(d). In that case, the current transport is contributed from both the surface and bulk of ITO channel. It is very interesting to note that when ITO-TFT is fully turned on, the channel can be actually regarded as a resistor with conductivity: $\sigma = q\mu N_D$. Thus, the channel current can be expressed by Ohm's law: $I_{DS} = q\mu N_D \left(t_{ITO} W/L\right) V_{DS}$, where $\mu$ is the channel carrier mobility and $N_D$ denotes carrier concentration of ITO film( $\sim 5\times 10^{19}/cm^3$). According to the above equation and transfer curve (see Fig.3(a) for $t_{ITO}$ =10 nm, $V_{DS}$=0.1V), the $\mu$ is estimated to be 8.8 cm$^2$/Vs.

For a traditional inversion-mode or accumulation-mode MOSFET, the channel has a high resistance to block the current through it under small gate voltages. In order to drive a significant channel current, a large gate voltage must be applied to create a thin conducting layer for carriers transport near dielectric/channel interface. As a result, the carrier scattering increases (and conductivity decreases) with increasing gate voltage. However, this does not occur in our junctionless ITO-TFTs where the current flows in the back side of the ITO (for the partially depleted case) or the entire ITO film (for the flat-band and accumulation cases). Thus, carrier transport in our devices is not influenced by the scattering centres at the dielectric/channel interface. The potential profile perpendicular to the gate layer (the *x* direction) can be understood through solving the 1D Poisson equation with a depletion approximation in the channel. With the relative permittivity of ITO, $\varepsilon_s \approx 9$ (ref. 1), spacial charge concentration of ITO film $\sim 5 \times 10^{19}$/cm$^3$ and a typical thickness of the EDL of $\sim 1.0$ nm[31-33], the depletion width of the ITO channel is thus estimated to be approximately 10 nm at $V_G = -2$ V. The calculated depletion width agrees well with our 10 nm thick ITO channel which can be fully switched off. The key points to fabricate such junctionless TFTs can be summarized as follows: (i) the ITO layer as a channel should be heavily doped to obtain a good ohmic contacts and a high on-state drain current; (ii) the channel layer should be sufficiently thin so that the charge carriers can be fully depleted at relatively small negative gate voltages; (iii) the $C_i$ should be as large as possible to achieve a strong field-effect modulation.

In order to check the repeatability of the device performance, the pulse response measurements are carried out, see Fig.5(a). It is seen that $I_{DS}$ repeats reasonably well in response to the pulsed $V_G$, suggesting that the on-state current and the on/off current ratio are highly repeatable. These results indicate that no chemical doping or chemical reaction occurs at the chitosan-based biopolymer electrolyte/ITO film interface. The electrostatic switching is dominant mechanism for our device operations. Here, we should acknowledge that our ITO-TFTs show a slow transient response due to the slow dielectric-polarizing limitation related to ion migration in the chitosan-based biopolymer electrolyte. However, for low-cost electronics applications, switching speed is not the primary parameter to achieve[34]. In this sense, the slow transient response may not be a killer to the low frequency applications such as sensors. In contrast, transparent feature and simple room-temperature process which is compatible with flexible electronics are the real merits for our devices[34]. They can find many potential applications in low-cost electronic devices[34]. To further investigate the device stability, a negative-bias-stress test is performed through applying a constant gate bias of $V_{GS}$= -2V. Figure 5(b) shows the evolution of the transfer curves as a function of the bias-stress time. The shift of $V_{th}$ ($\Delta V_{th}$) increases with the stress time. From this figure, the $\Delta V_{th}$ is found to be ~0.43V for the 1000s-stress curve. In general, the origins of $V_{th}$ shift are attributed to charge trapping at the gate dielectric and/or the interface between the channel and gate dielectric, or else resulted from the defects in the channel[35,36]. For our junctionless ITO-TFTs, the $V_{th}$ shift could originate from (i) $O_2$ (from ambient atmosphere) related charge trapping centres and (ii) the structural defects induced during depositing the ITO channel layer. As the ITO layer is very thin, $O_2$ molecules could possibly penetrate it into the chitosan film. Direct ion bombardment onto chitosan in the subsequent ITO sputtering deposition can also induce structural defects at the surface of chitosan and finally degrade the device performance.

To eliminate these possible factors, a 5-nm thick $SiO_2$ film is coated on top of chitosan layer to form an organic/inorganic hybrid bilayer. Figure 6 (a)~(d) show the electrical characteristics of the junctionless ITO-TFTs with such a hybrid bilayer dielectric. Compared to the ITO-TFT with a pure chitosan gate dielectric, the TFT with the reinforced hybrid dielectric exhibits an improved performance, i.e., a larger $I_{on/off}$ of $5.5 \times 10^7$ and a smaller $S$ of 84mV/dec, see Fig 6(a). In addition, the $\mu$ exhibits a large value of 23 $cm^2/Vs$ according to the Ohm's law of bulk current in ITO channel. The device also exhibits good Ohmic contact properties as good current saturation behaviors at high $V_{DS}$ and linear characteristics at low $V_{DS}$ are observed, see Fig. 6(b). The pulse response of the junctionless TFTs gated by the hybrid bilayer dielectric with $t_{ITO}$=10nm is improved in comparison with the device with pure chitosan electrolyte, as $I_{on/off}$ maintains a larger value >$10^7$, see Fig. 6(c). Again, $\Delta V_{th}$ is found to be only 0.13V for the 1000s-stress curve, a factor of three smaller than that for the devices with the pure chotosan dielectric, as shown in Fig. 6(d).

Another feature of our ITO-FETs is highly transparent. Figure 7 shows the optical transmission spectra of the junctionless organic-inorganic hybrid ITO-TFTs in the wavelength range between 200 and 1000 nm. The average transmittance in the visible portion (400–700 nm) is more than 80%. A photograph of the ITO-TFT arrays placed on a background text is shown in the inset of Fig.7. One can see the text clearly through the devices, while the junctionless TFT arrays are invisible.

In summary, junctionless transparent organic-inorganic hybrid ITO thin-film transistors with a very simple structure have been fabricated using very simple fabrication processes. The channel and source/drain electrodes of the ITO-TFTs are realized through a heavily-doped ITO-coplanar ultra-thin film. Solution-processed chitosan-based biopolymer electrolyte and a reinforced chitosan/SiO$_2$ hybrid bilayer dielectric stack are employed respectively as gate dielectric. We find that a 10 nm thick ITO film with a reinforced chitosan/SiO$_2$ hybrid dielectric exhibits an excellent FET performance with a small subthreshold swing of 84mV/dec and a large on/off ratio of $5.5 \times 10^7$. Such a simple device structure and fabrication process may open the door for easy access to low-cost electronic applications.

*Acknowledgments:*

**The authors are grateful for the financial supports from a Foundation for the Author of National Excellent Doctoral Dissertation of P R China (Grant No. 200752), and Fok Ying Tung Education Foundation (Grant No. 121063)**

# References

1 Opnescu, A. M. Nanowire transistors made easy. *Nat. Nanotechnol*. **5**, 178-179 (2010).

2 Lilienfeld, J. E. Method and apparatus for controlling electric current. US patent 1, 745, 175 (1925).

3 Colinge, J. P., Lee, C. W., Afzalian, A., Akhavan, N. D., Yan, R., Ferain, I., Razavi, P., Neill, B., Blake, A., White, M., Kelleher, A. M., McCarthy, B. & Murphy, R. Nanowire transistors without junctions. *Nat. Nanotechnol.* **5,** 225-229 (2010).

4 Colinge, J. P., Lee, C. W., Ferain, I., Akhavan, N. D., Yan, R., Razavi, P., Yu, R., Nazarov, A. N. & Doriac, R. T. Reduced electric field in junctionless transistors. *Appl. Phys. Lett.* **96,** 073510 (2010).

5 Gundapaneni, S., Ganguly, S. & Kottantharayil, A. Bulk planar junctionless transistor (BPJLT): an attractive device alternative for scaling. *IEEE Electron Device Lett*. **32,** 261-263 (2011).

6 Lin, H. -C., Lin, C. -I. & Huang, T. -Y. Characteristics of n-type junctionless poly-Si thin-film transistors with an ultrathin channel. *IEEE Electron Device Lett*. **33,** 53-55 (2012).

7 Choi, S. -J., Moon, D. -I., Kim, S., Ahn, J. -H., Lee, J. -S., Kim, J. -Y. & Choi, Y. -K. Nonvolatile memory by all-around-gate junctionless transistor composed of silicon nanowire on bulk substrate. *IEEE Electron Device Lett*. **32,** 602-604 (2011).

8 Nomura, K., Ohta, H., Takagi, A., Kamiya, T., Hirano, M. & Hosono, H. Room-temperature fabrication of transparent flexible thin-film transistors using amorphous oxide semiconductors. *Nature* **432,** 488-492 (2004).

9 Fortunato, E., Barquinha, P., Pimentel, A., Goncalves, A., Marques, A., Pereira, L. & Martins, R. Fully transparent ZnO thin-film transistor produced at room temperature. *Adv. Mater.* **17,** 590-594 (2005).

10 Lim, W., Jang, J. H., Kim, S.-H., Norton, D. P., Craciun, V., Pearton, S. J., Ren, F. & Shen, H. High performance indium gallium zinc oxide thin-film transistors fabricated on polyethylene terephthalate substrates. *Appl. Phys. Lett*. **93,** 082102 (2008).

11 Tahar, R. B. H., Ban, T., Ohya, Y. & Takahashi, Y. Tin doped indium oxide thin films: electrical properties. *J. Appl. Phys*. **83,** 2631 (1998).

12 Wan, Q., Dattoli, E. N., Fung, W. Y., Guo, W., Chen, Y. B., Pan, X. Q. & Lu, W. High-performance transparent conducting oxide nanowires. *Nano Lett*. **6,** 2909-2915 (2006).

13 Yoon, M. -H., Yan, H., Facchetti, A. & Marks, T. J. Low-voltage organic field-effect transistors and inverters enabled by ultrathin cross-linked polymers as gate dielectrics. *J. Am. Chem. Soc.* **127,** 10388-10395 (2005).

14 DiBenedetto, S. A., Paci, I., Facchetti, A., Marks, T. J. & Ratner, M. A. High-capacitance organic nanodielectrics: effective medium models of their response. *J. Phys. Chem. B*. **110,** 22394-22399 (2006).

15 Ha, Y., Jeong, S., Wu, J., Kim, M. -G., Dravid, V. P., Facchetti, A. & Marks, T. J. Flexible low-voltage organic thin-film transistors enabled by low-temperature, ambient solution-processable inorganic/organic hybrid gate dielectrics. *J. Am. Chem. Soc.* **132,** 17426-17434 (2010).

16 Ha, Y., Emery, J. D., Bedzyk, M. J., Usta, H., Facchetti, A. & Marks, T. J. Solution-deposited organic-inorganic hybrid multilayer gate dielectrics. design, synthesis, microstructures, and electrical properties with thin-film transistor. *J. Am. Chem. Soc.* **133,** 10239-10250 (2011).

17 Cho, J. H., Lee, J., Xia, Y., Kim, B., He, Y., Renn, M. J., Lodge, T. P. & Frisbie, C. D. Printable ion-gel gate dielectrics for low-voltage polymer thin-film transistors on plastic. *Nature Mater.* **7,** 900-906 (2008).

18 Panzer, M. J. & Frisbie, C. D. Polymer electrolyte-gated organic field-effect transistors: low-voltage, high-current switches for organic electronics and testbeds for probing electrical transport at high charge carrier density. *J. Am. Chem. Soc.* **129,** 6599-6607 (2007).

19 Braga, D., Ha, M. J., Xie, W. & Frisbie, C. D. Ultralow contact resistance in electrolyte-gated organic thin film transistors. *Appl. Phys. Lett.* **97,** 193311 (2010).

20 Herlogsson, L., Crispin, X., Robinson, N. D., Sandberg, M., Hagel, O.-J., Gustafsson, G. & Berggren, M. Low-voltage polymer field-Effect transistors gated via a proton conductor. *Adv. Mater.* **19,** 97-101 (2007).

21 Malti, A., Gabrielsson, E. O., Berggren, M. & Crispin, X. Ultra-low voltage air-stable polyelectrolyte gated n-type organic thin film transistors. *Appl. Phys. Lett.* **99,** 063305 (2011).

22 Yuan, H., Shimotani, H., Tsukazaki, A., Ohtomo, A., Kawasaki, M. & Iwasa, Y. High-density carrier accumulation in ZnO field-effect transistors gated by electric double layers of ionic liquids. *Adv. Funct. Mater.* **19,** 1046-1053 (2009).

23 Jiang, J., Wan, Q., Sun, J. & Lu, A. Vertical low-voltage oxide transistors gated by microporous $SiO_2$/LiCl composite solid electrolyte with enhanced electric-double-layer capacitance. *Appl. Phys. Lett.* **97,** 052104 (2010).

24 Jiang, J., Sun, J., Zhou, B., Lu, A. & Wan, Q. Self-assembled in-plane gate oxide-based homojunction thin-film transistors. *IEEE Electron Device Lett.* **32,** 500-502 (2011).

25 Jiang, J., Sun, J., Dou, W., Zhou, B. & Wan, Q. Junctionless in-plane-gate transparent thin-film transistors. *Appl. Phys. Lett.* **99,** 193502 (2011).

26 Jiang, J., Sun, J., Zhu, L., Wu, G. & Wan, Q. Dual in-plane-gate oxide-based thin-film transistors with tunable threshold voltage. *Appl. Phys. Lett.* **99,** 113504 (2011).

27 Fernandes, S., Freire, C., Silvestre, A., Neto, C. & Gandini, A. Novel materials based on chitosan and cellulose. *Polym. Int.* **60,** 875-882 (2011).

28 Palla, C., Pacheco, C. & Carrin, M. Preparation and modification of chitosan particles for Rhizomucor miehei lipase immobilization. *Biochem. Eng. J.* **55,** 199-207 (2011).

29 Larsson, O., Said, E., Berggren, M. & Crispin, X. Insulator polarization mechanisms in polyelectrolyte-gated organic field-effect transistors. *Adv. Funct. Mater.* **19,** 3334-3341 (2009).

30 Lee, J., Kaake, L., Cho, J. H., Zhu, X.-Y., Lodge, T. P. & Frisbie, C. D. Ion gel-gated polymer thin-film transistors: operating mechanism and characterization of gate dielectric capacitance, switching speed, and stability. *J. Phys. Chem. C*. **113,** 8972-8981 (2009).

31 Uemura, T., Hirahara, R., Tominari, Y., Ono, S., Seki, S. & Takeya, J. Electronic functionalization of solid-to-liquid interfaces between organic semiconductors and ionic liquids: Realization of very high performance organic single-crystal transistors. *Appl. Phys. Lett.* **93,** 263305 (2008).

32 Takeya, J., Yamada, K., Hara, K., Shigeto, K., Tsukagoshi, K., Ikehata, S. & Aoyagi, Y. High-density electrostatic carrier doping in organic single-crystal transistors with polymer gel electrolyte. *Appl. Phys. Lett.* **88,** 112102 (2006).

33 Ono, S., Miwa, K., Seki, S. & Takeya, J. A comparative study of organic single-crystal transistors gated with various ionic-liquid electrolytes. *Appl. Phys. Lett*. **94,** 063301 (2009).

34 Facchetti, A. Dielectric materials: gels excel. *Nature Materials*. **7,** 839-840 (2008).

35 Park, K.-B., Seon, J.-B., Kim, G. H., Yang, M., Koo, B., Kim, H. J., Ryu, M.-K. & Lee, S.-Y. High electrical performance of wet-processed indium zinc oxide thin-film transistors. *IEEE Electron Device Lett*. **31,** 311-313 (2010).

36 Kim, K. M., Kim, C. W., Heo, J. -S., Na, H., Lee, J. E., Park, C. B., Bae, J.-U., Kim, C. -D., Jun, M., Hwang, Y. K., Meyers, S. T., Grenville, A. & Keszler, D. A. Competitive device performance of low-temperature and all-solution-processed metal-oxide thin-film transistors. *Appl. Phys. Lett.* **99,** 242109 (2011).

**Figure captions**

Fig.1. The schematic diagram of our junctionless chitosan/SiO$_2$ hybrid ITO-TFT.

Fig.2. (a) Specific gate capacitance of the junctionless TFTs without and with the SiO$_2$ layer in the frequency range from 20Hz to 1M Hz. The inset shows the molecular structures of chitosan. (b) Proton hopping mechanism: when an electric field is applied, some of the protonated amino groups will be deprotonated and then transport through hopping from one oxygen atom to another with hydrogen bonds. (c) Gate leakage curves of the ITO-TFTs without and with the SiO$_2$ layer.

Fig.3. (a) The transfer curves of the ITO-TFTs gated through pure chitosan-based biopolymer dielectric with several different thick ITO layers ($t_{ITO}$=10nm, 20nm, 30nm, 60nm). (b) The output curves of the ITO-TFTs for $t_{ITO}$= 10nm.

Fig.4. The energy band diagrams and schematic diagrams for the ITO-FETs at four working modes. (a) Fully depleted mode ($V_G<<0$). (b) Partly-depleted channel mode ($V_G<0$). (c) Neutral channel mode ($V_G \approx 0$). (d) Surface-accumulated channel mode ($V_G > 0$).

Fig.5. The ITO-TFTs with pure chitosan as the gate dielectric for $t_{ITO}$= 10nm. (a) Pulse response with a square-shaped $V_G$ of $V_+$= 2.0V and $V_-$= -2.0V under a constant bias of $V_{DS}$=2.0V. (b) The evolution of the $I_{DS}$-$V_{GS}$ curves of the ITO-TFT to which a negative-stress bias ($V_{GS}$= -2V) is applied to the device as a function of the stress time.

Fig.6. The ITO-TFTs with chitosan/SiO$_2$ hybrid bilayer as the gate dielectric for $t_{ITO}$= 10nm. (a) The $I_{DS}$-$V_{GS}$ curves. (b) The $I_{DS}$-$V_{DS}$ curve. (c) Pulse response with a square-shaped $V_G$ of $V_+$= 2.0V and $V_-$= -2.0V under a constant bias of $V_{DS}$=2.0V. (d)

The evolution of the $I_{DS}$-$V_{GS}$ curves of the device to which a negative-stress bias($V_{GS}$= -2V) is applied as a function of the stress time.

Fig.7 Optical transmittance spectrum of the hybrid ITO-TFTs on glass substrate. The inset shows an optical image of an as-prepared hybrid ITO-TFT array on a glass substrate placed on a text paper.

Jiang Jie *et al.*
The gated conductance
of thin indium tin oxide

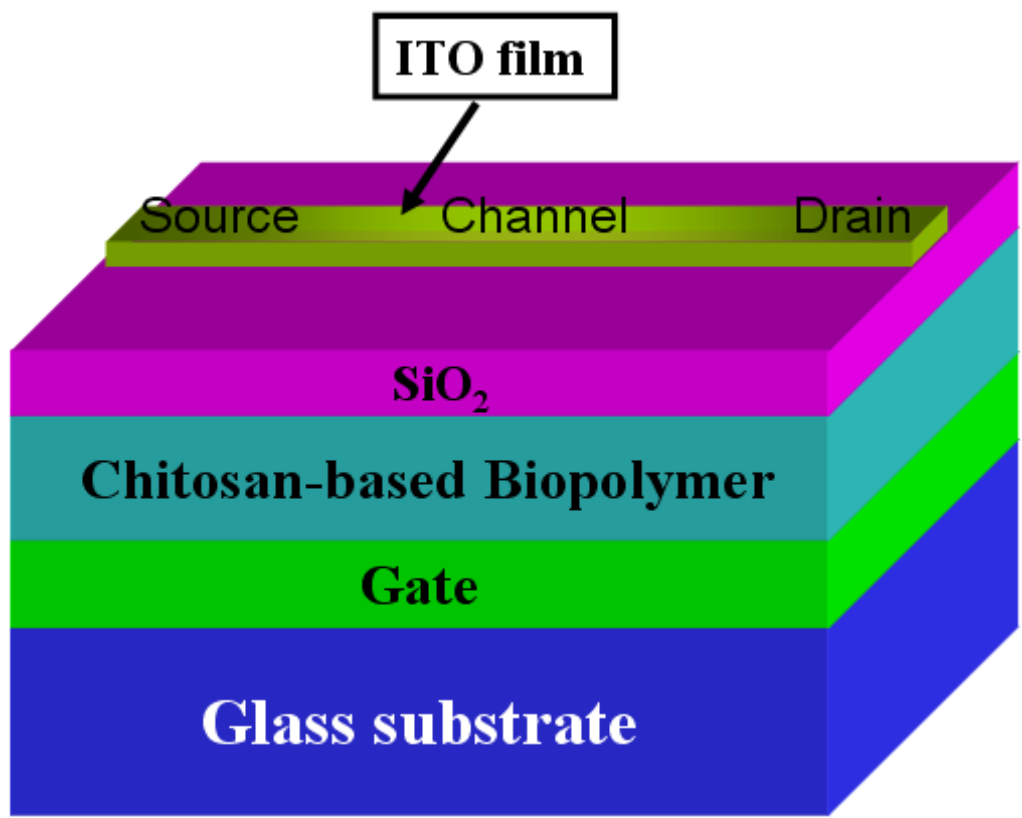

Figure 1

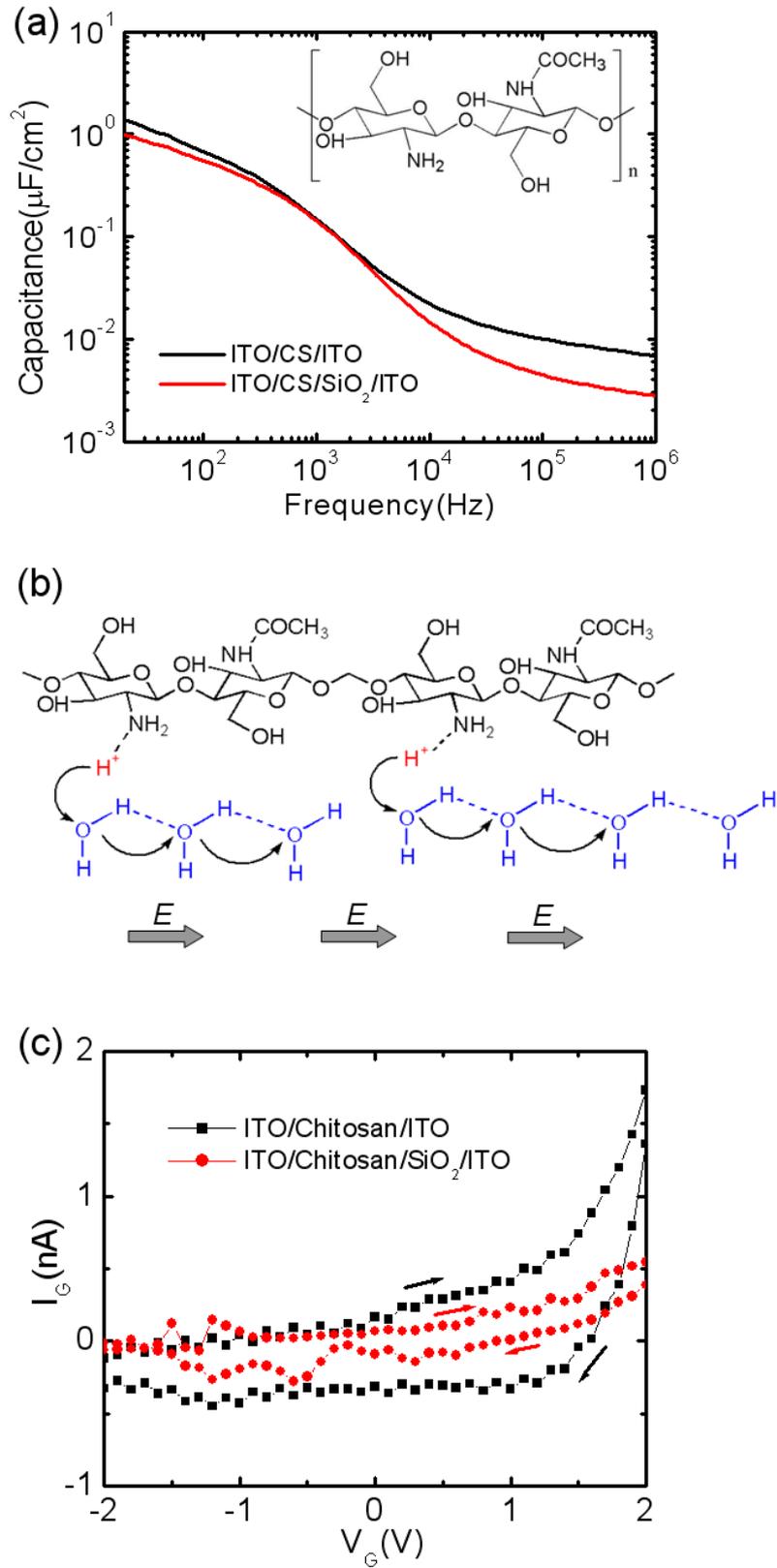

Figure 2

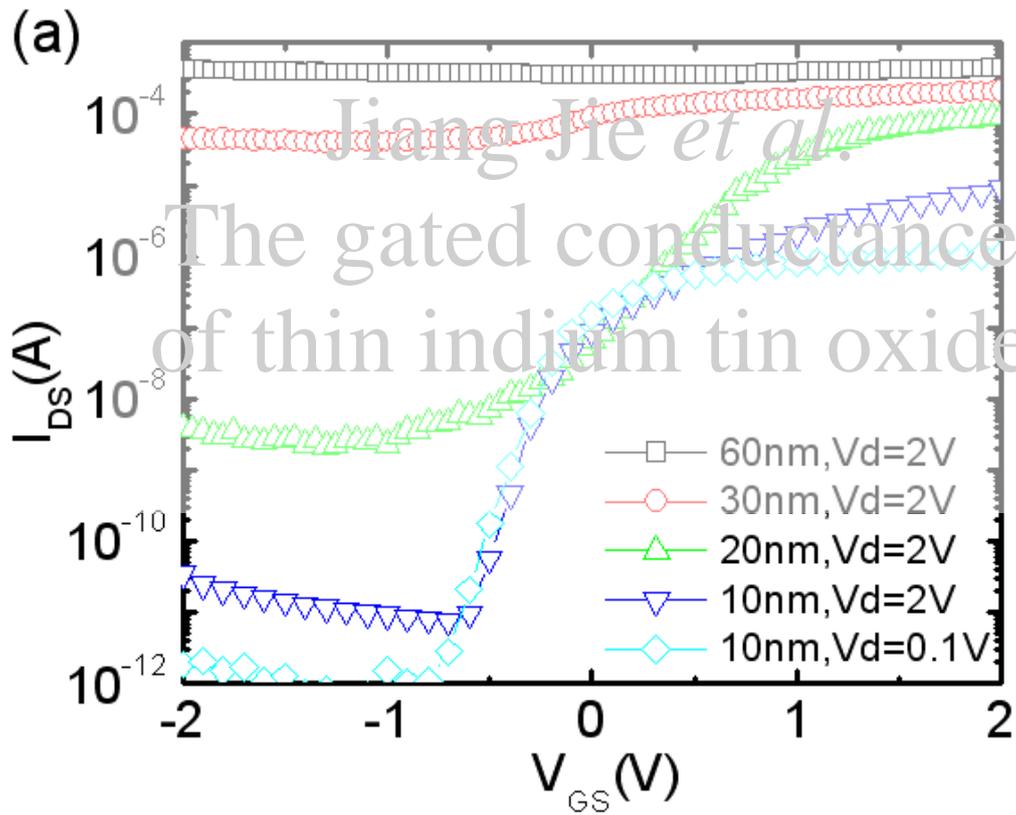
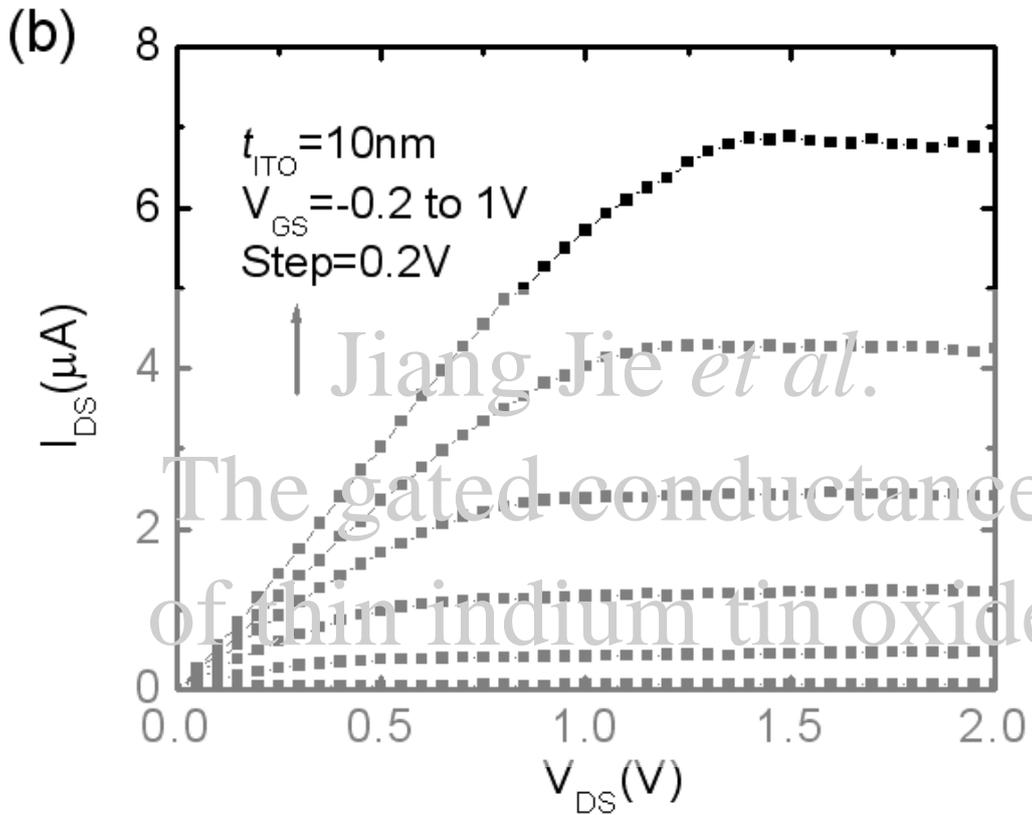

Figure 3

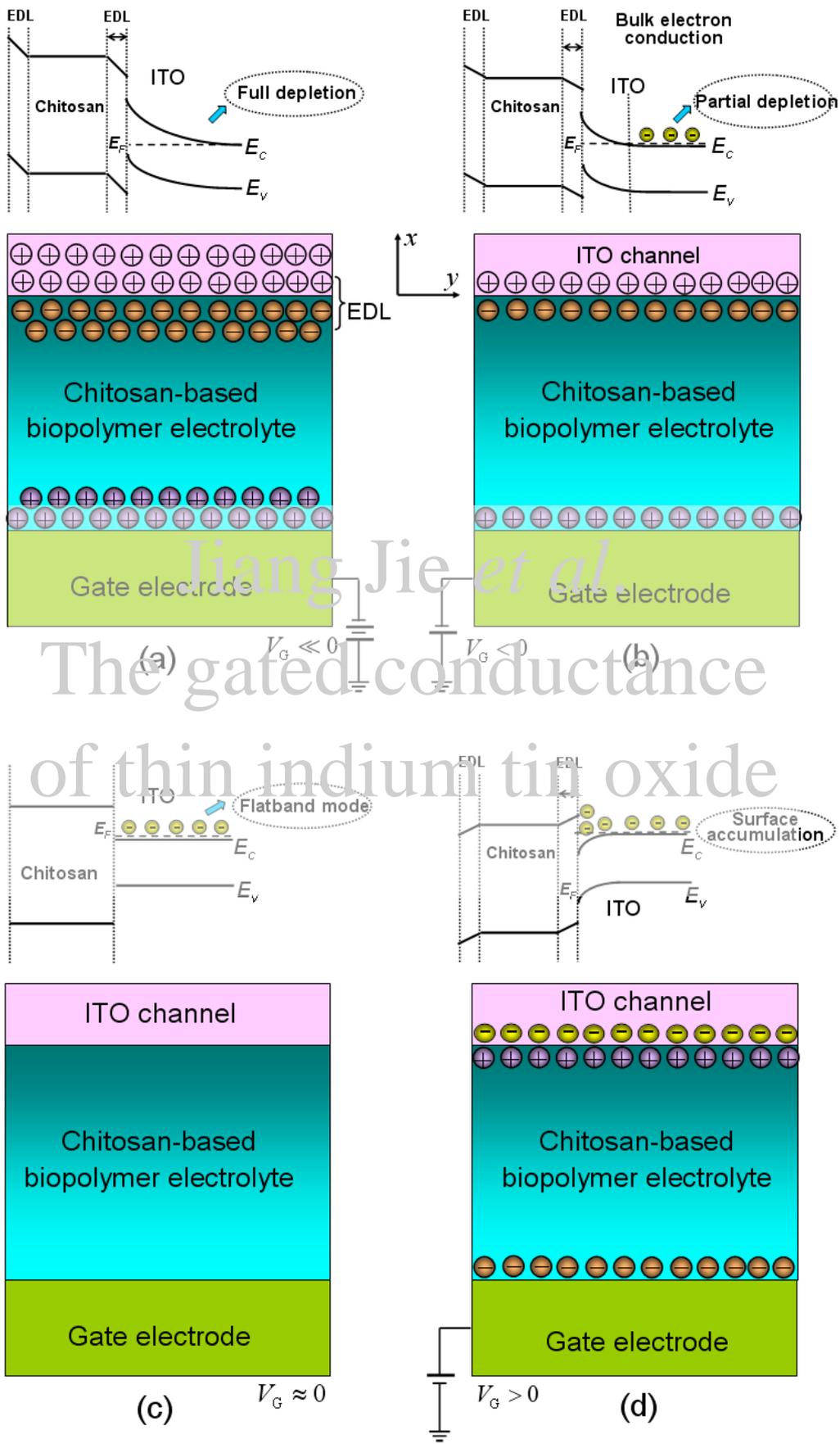

Figure 4

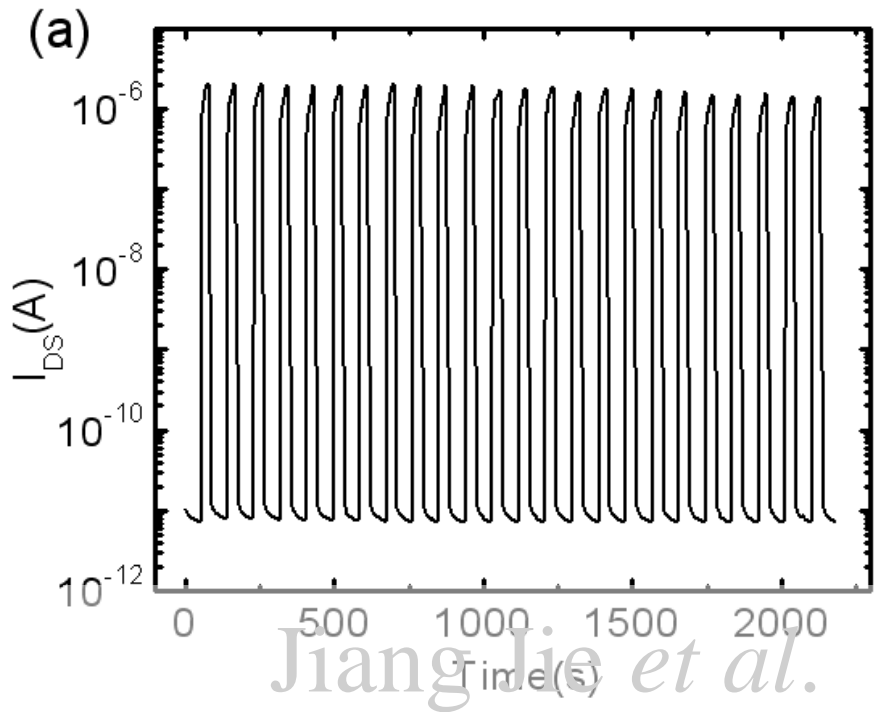

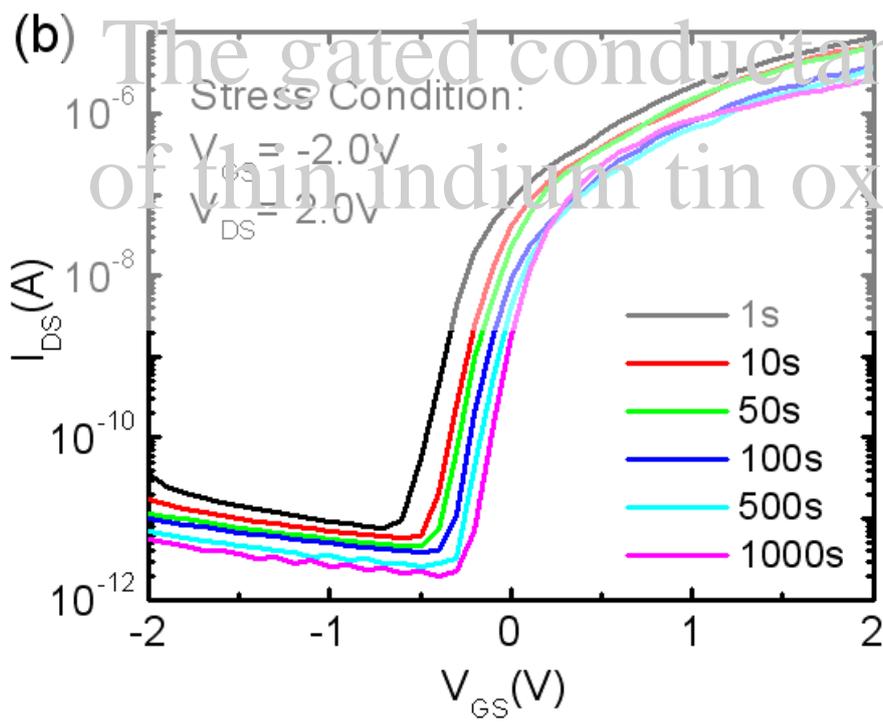

Figure 5

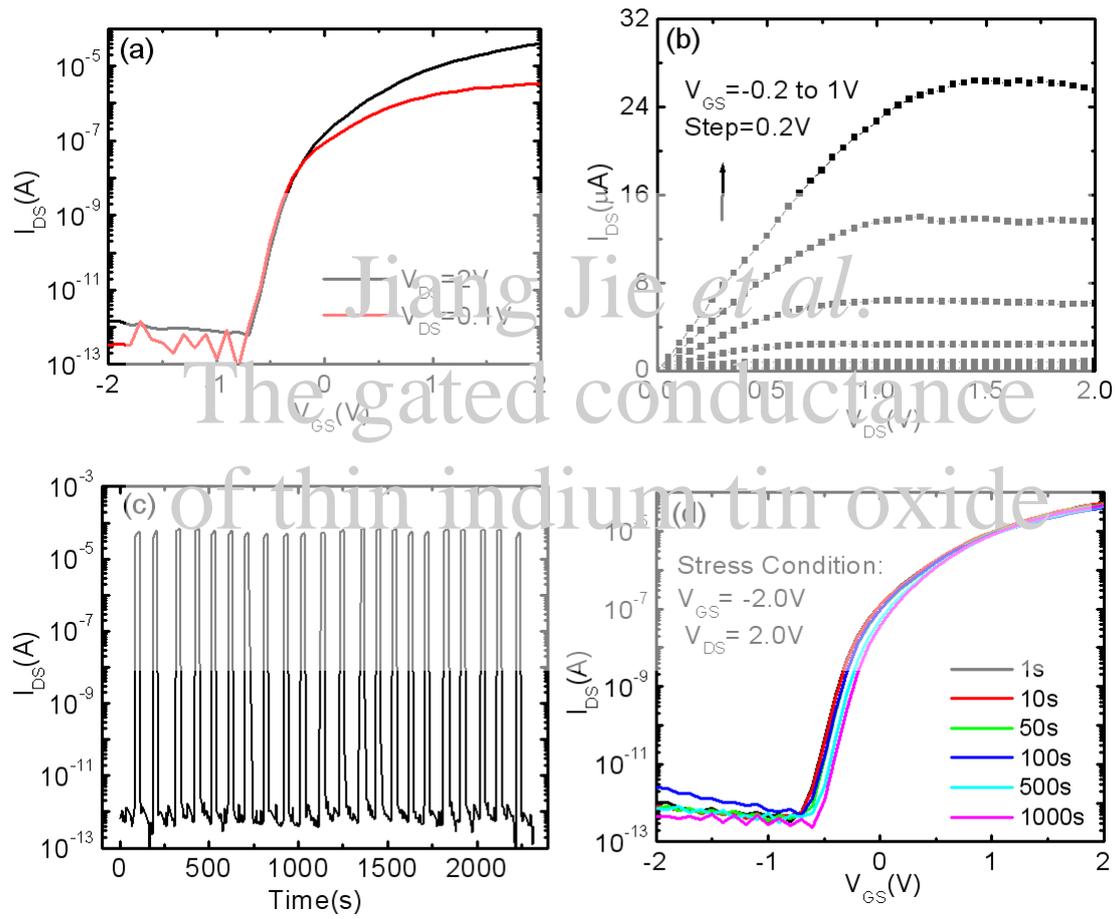

Figure 6

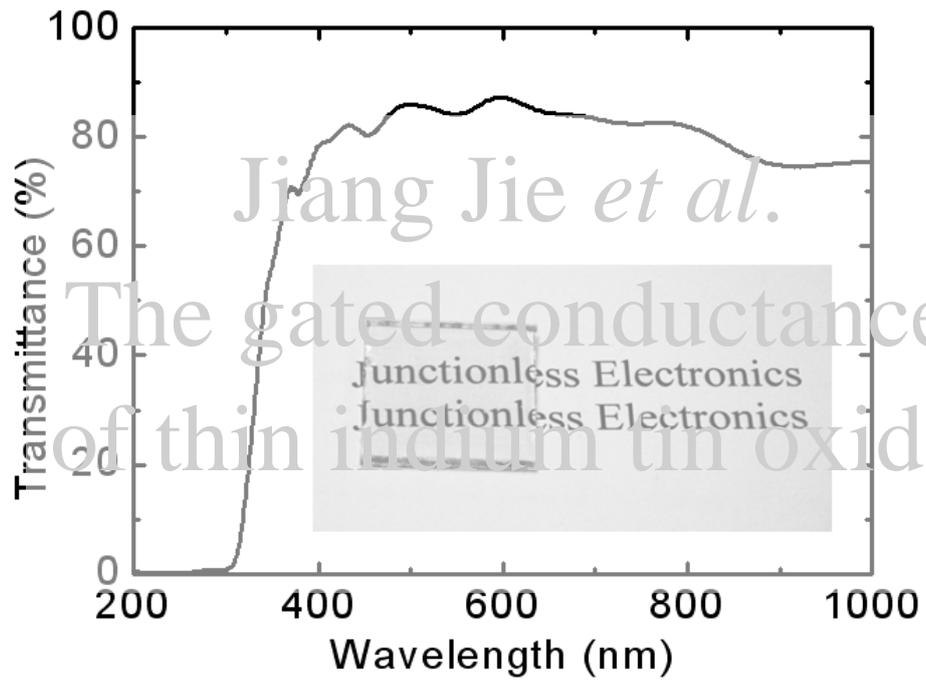

Figure 7